\documentclass[12pt]{article}

\usepackage{scicite}

\usepackage{times}

\usepackage{amsmath}
\usepackage{amsfonts}
\usepackage{amssymb}
\usepackage[usenames, dvipsnames]{color}
\usepackage{graphicx}
\usepackage{subfigure}
\usepackage{ulem}

\newenvironment{sciabstract}{%
\begin{quote} \bf}
{\end{quote}}

\newcommand{\fig}[1]{Fig.~\ref{#1}}
\newcommand{\eq}[1]{Eq.~\ref{#1}}

\renewcommand{\vec}[1]{\boldsymbol{#1}}
\newcommand{\tens}[1]{\boldsymbol{#1}}
\newcommand{\bnabla}{\vec{\nabla}}

\title{Onset of meso-scale turbulence in living fluids}

\author{Amin Doostmohammadi,${}^{1\ast}$
	Tyler N. Shendruk,${}^{1\ast}$
	Kristian Thijssen,${}^{2\ast}$\\
	Julia M. Yeomans${}^{1}$
\\
\normalsize{${}^{1}$The Rudolf Peierls Centre for Theoretical Physics,}\\
\normalsize{1 Keble Road, Oxford, OX1 3NP, United Kingdom}\\
\normalsize{${}^{2}$Department of Applied Physics, Eindhoven University of Technology,}\\
\normalsize{5600 MB, Eindhoven, Netherlands}\\
\\
~\\
\normalsize{$^\ast$These authors contributed equally to this work}\\
}
\date{}

\begin{document} 


\baselineskip24pt


\maketitle 



\begin{sciabstract}
Meso-scale turbulence is an innate phenomenon, distinct from inertial turbulence, that spontaneously occurs at {\em zero}-Reynolds number in fluidized biological systems.
This spatio-temporal disordered flow radically changes nutrient and molecular transport in living fluids and can strongly affect the collective behaviour in prominent biological processes, including biofilm formation, morphogenesis and cancer invasion. 
Despite its crucial role in such physiological processes, understanding meso-scale turbulence and any relation to classical inertial turbulence remains obscure.
Here, we show how the motion of active matter along a micro-channel transitions to meso-scale turbulence through the evolution of disordered patches (active puffs) from an absorbing state of flow vortex-lattices.
We demonstrate that the critical behaviour of this transition to meso-scale turbulence in a channel belongs to the directed percolation universality class.
This finding bridges our understanding of the onset of zero-Reynolds number meso-scale turbulence and traditional scale-invariant turbulence, therefore generalizing theories on the onset of turbulence in confinement to the distinct classes of incoherent flows observed in biological fluids.
\end{sciabstract}

Despite extensive implications for diverse fluid dynamical systems and more than a century of research, the transition from pressure-driven laminar flow to inertial turbulence in even the simplest geometries remains one of the major unresolved problems in fluid mechanics. Elaborate experiments have recently shed new light on the nature of this transition by measuring the decay and splitting of local turbulent domains/clusters (puffs) in pipe flows and have determined the critical Reynolds number -- ratio of inertial to viscous forces -- at which the transition occurs \cite{Hof2011,Hof2015}. Short-range interactions between the locally turbulent puffs, which feed on surrounding laminar flow as an absorbing state, drive a continuous transition to a fully turbulent flow. Recent experimental evidence from channel and circular Couette flows \cite{Sano2015,Hof2016} together with direct numerical simulation studies and predator-prey models \cite{Shih2015}, have provided evidence that the transition at the critical Reynolds number is characterised by the directed percolation universality class. 
Strikingly, here we show that for a profoundly distinct class of turbulence at {\em zero}-Reynolds number, the transition in a channel can also be characterised by the emergence of spontaneous puffs created by microscopic activity of biological fluids. Even for this zero-Reynolds number class of turbulent-like flows, we find that the critical exponents belong to the directed percolation universality class.

Zero-Reynolds number turbulence is established through continuous energy injection from the constituent elements of an active fluid in many biological systems, including bacterial suspensions \cite{Dombrowski2004,Julia2012,Aranson2012,Aranson2014,Sano2016bacteria,Wioland2016}, cellular monolayers \cite{Benoit2012,ourSM2015,Sano2016cells}, or sub-cellular filament/motor protein mixtures \cite{Dogic2012,Francesc2016}.
Although the inertia is negligible (Reynolds number $\sim 10^{-6}$) in such systems, active turbulence is characterised by a highly disordered distribution of vortices \cite{Frey2015,Giomi2015}. 
However, meso-scale turbulence in living fluids possesses a characteristic vortex length scale, which distinguishes it from scale-invariant inertial turbulence \cite{Heidenreich2015}, and it is considered a new class of turbulent flow \cite{Julia2012,Jorn2013,Frey2015}. 

\begin{figure}
\centering
\includegraphics[trim = 325 100 0 0, clip, width=1.0\textwidth]{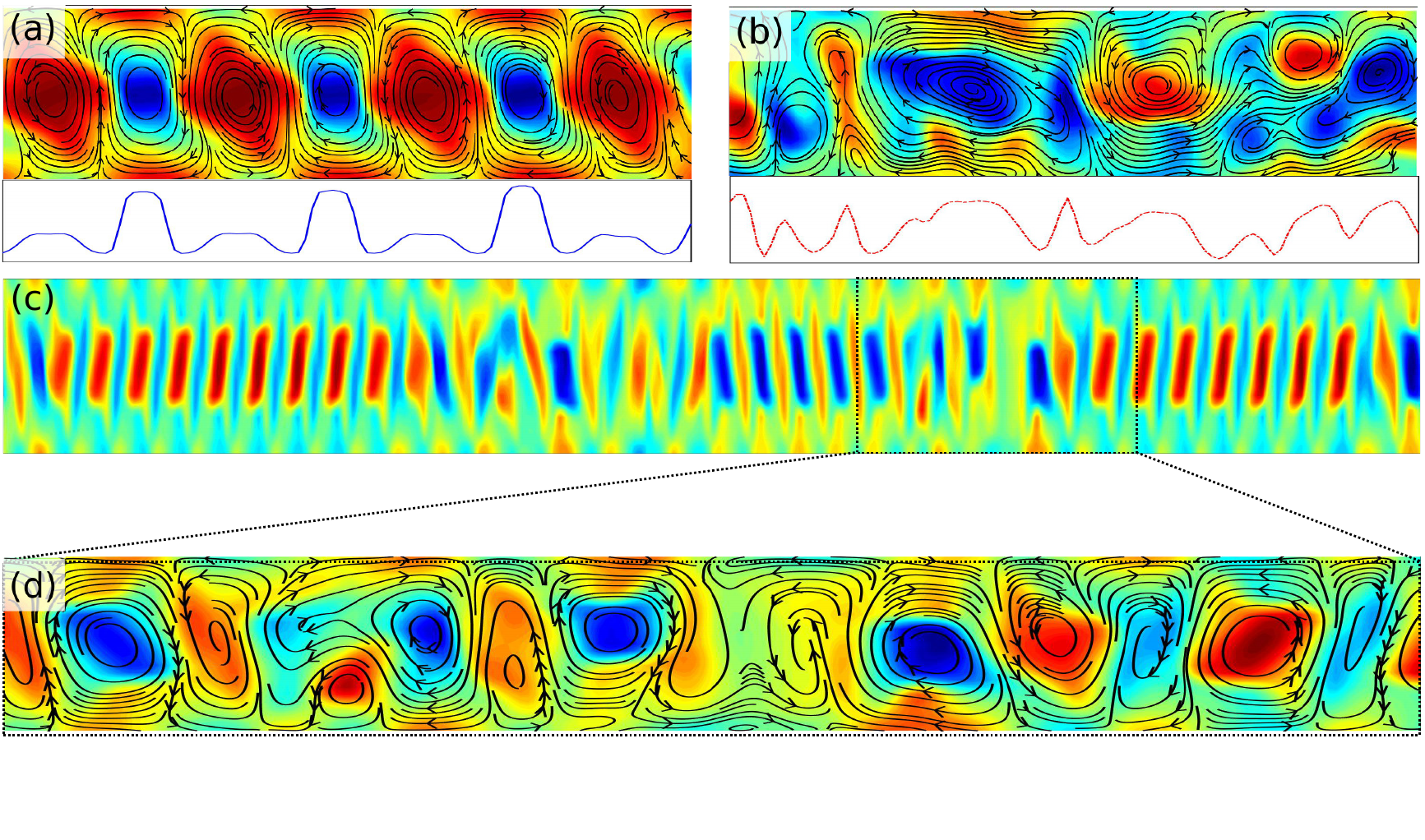}
\caption{{\bf Emergence of active puffs from a vortex-lattice derives the transition to turbulence in living fluids.} (a) A highly ordered flow vortex-lattice is formed at lower activities and (b) active turbulence is fully established at higher activities. Lower panels in (a), (b) show the height-averaged enstrophy signal along the channel. (c) Coexistence of the vortex-lattice and meso-scale turbulence close to the transition point.
~The zoomed-in panel in (d) illustrates the spontaneous formation of active puffs from the vortex-lattice. Colormaps show vorticity contours with blue and red colors corresponding to clockwise and anti-clockwise vortices, respectively. Solid black lines illustrate streamlines of the flow.}
\label{fig:flows}
\end{figure}

\begin{figure}
\centering
\includegraphics[width=1.0\textwidth]{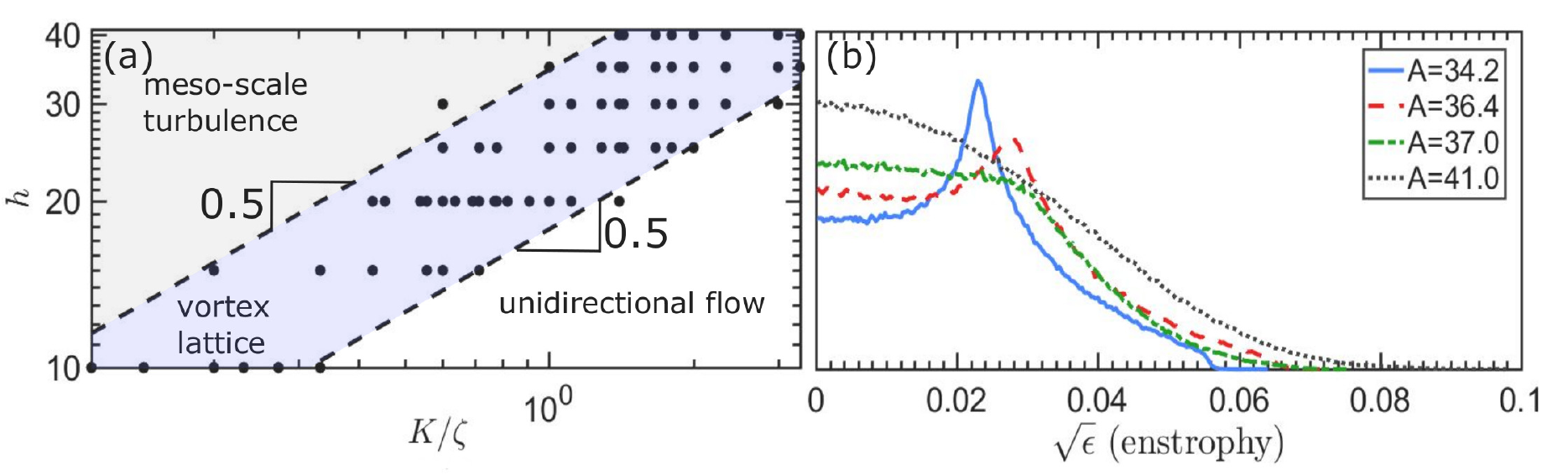}
\caption{{\bf Competing length scales control the transition to active turbulence.} (a) Phase-space of control parameters corresponding to the vortex-lattice state. The slope of $0.5$ at the boundaries of the vortex-lattice state shows that the emergence of vortex-lattice is controlled by two competing length scales: the channel height $h$ and the activity length scale $\sqrt{K/\zeta}$. (b) Transition from the vortex-lattice state to active turbulence is characterised by the channel-averaged enstrophy $\epsilon$ distribution for increasing values of activity number $A=\sqrt{\zeta h^{2}/K}$.}
\label{fig:contP}
\end{figure}

To study the transition to meso-scale turbulence, we computationally solve the continuum equations of active nematics in micro-channels, which have successfully reproduced the patterns of bacterial ordering in bulk \cite{Sano2016bacteria} and in confinement \cite{Volfson2008}, the flow structure and correlation lengths of microtuble bundles \cite{Dogic2012,ourprl2013,Francesc2016} and the flow patterns of dividing cells \cite{ourSM2015} (see Supplementary Information for the details of the model). 
In this zero-Reynolds number regime, the transition to turbulence occurs by increasing the amount of local energy injection (activity) in the living fluids. In a confined environment, the activity leads to spontaneous symmetry breaking and the generation of unidirectional flow \cite{Joanny2005}, which is followed by an oscillatory regime characterised by distorted streamlines \cite{Giomi2012,Julicher2016}, upon increasing the activity. Further increase in the activity leads to the emergence of a stable lattice of velocity vortices throughout the channel [\fig{fig:flows}(a)], and this transitions to meso-scale turbulence at higher activities [\fig{fig:flows}(b)]. The emergence of the intermediate vortex-lattice in active matter has been observed experimentally in motility assays of microtubles \cite{Chate2012}, in bacterial suspension in a channel confinement \cite{Wioland2016}, and also numerically by hydrodynamic screening of activity-induced flows due to frictional damping \cite{ourNComm2016}. In stark contrast to inertial turbulence, the Reynolds number is irrelevant here and the transition between flow regimes is governed by the dimensionless {\it activity number} $\mathrm{A} = \sqrt{\zeta h^2/K}$ [\fig{fig:contP}(a)]. This parameter characterises the ratio of the channel height $h$, which here is equivalent to the hydrodynamic screening length, to the characteristic activity-induced length scale $\ell_{a}=\sqrt{K/\zeta}$, that represents the relative importance of the intrinsic activity $\zeta$ and the orientational elasticity $K$ of the nematic fluid. 

The marked difference between the various flow states is clearly seen in the structure of the vorticity. Therefore, in order to characterise the transition between the regimes, we measure the distribution of the local enstrophy $\epsilon$, averaged across the channel. This quantity represents the strength of vortices in the flow, and has also been used for determining the nature of inertial turbulence. The vortex-lattice state possesses a well defined peak in the enstrophy [\fig{fig:contP}(b)]. As the active flow transitions at higher activities, the enstrophy distribution broadens demonstrating that vorticity cascades down into meso-scale turbulence. 
The gradual disappearance of the peak in the enstrophy distribution [\fig{fig:contP}(b)] suggests a continuous transition from the vortex-lattice to meso-scale turbulence. But how does the active turbulence develop from the vortex-lattice?

Figure \ref{fig:flows}(c) shows a snapshot of the vorticity field in a long channel close to the transition. The vortex-lattice predominantly occupies the entire channel. Locally, however, we can identify regions of the channel where vortex pairs split into smaller non-ordered vortices [\fig{fig:flows}(d)]. This coexistence of the global vortex-lattice and clusters of local active turbulence controls the transition to turbulence in the channel. We term these localized domains of non-ordered vorticity {\it active puffs}, in analogy to the inertial puffs observed in the experiments on scale-invariant turbulence in long tubes \cite{Hof2011}. Unlike the inertial puffs that are externally initiated by perturbations to the flow field (such as induced pressure jumps), active puffs arise spontaneously due to the innate active forcing of the flow.

\begin{figure}
\centering
\subfigure[]
{\includegraphics[width=0.301\textwidth]{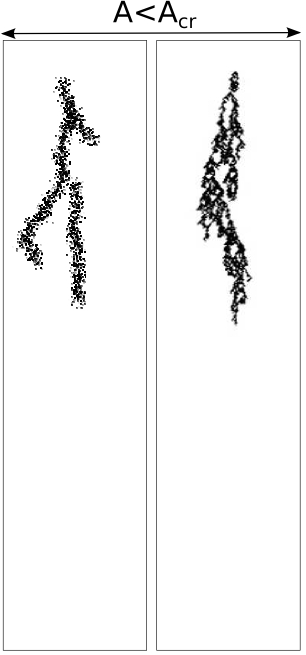}}
\subfigure[]
{\includegraphics[width=0.298\textwidth]{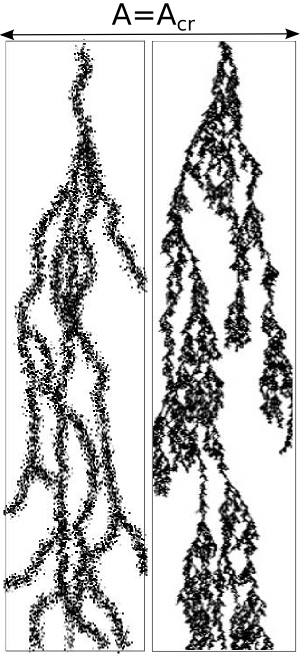}}
\subfigure[]
{\includegraphics[width=0.3008\textwidth]{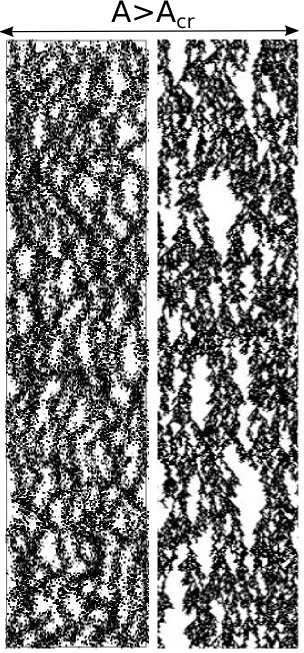}}
\caption{{\bf Active turbulence percolates over time as the active puffs split and decay}. Spatio-temporal evolution of active puffs represented by space-time kymograph of the height-averaged enstrophy (a) below the critical activity $A<A_{cr}$, (b) at the critical activity $A=A_{cr}$, and (c) above the critical activity $A>A_{cr}$. In (a), (b), (c) left panels correspond to active turbulence and right panels show simulations from the directed percolation model.}
\label{fig:puff}
\end{figure}

The emergence and dynamics of active puffs is clearly characterised in the space-time kymograph of enstrophy [\fig{fig:puff}]. At initial times, the entire channel is in the absorbing vortex-lattice state. However, depending on the activity number, active puffs spontaneously spread through the channel length [\fig{fig:puff},~Fig.~S.1]. A puff can split, giving birth to new puffs, while for every given moment there is a finite chance that an active puff decays back to the absorbing state [\fig{fig:puff}(a)] or that a new puff is spontaneously born. Above some critical activity number, this competition between decaying to the absorbing state and splitting produces a statistical steady-state in which active puffs coexist with the vortex-lattice. The coexistence results in a well-defined turbulence fraction within the channel [\fig{fig:puff}(b)]. At the highest activities, the active flow approaches the fully turbulent state, with active puffs ultimately occupying the entire channel [\fig{fig:puff}(c)]. 

\begin{figure}
\centering
\includegraphics[trim = 0 500 500 0, clip, width=1.0\textwidth]{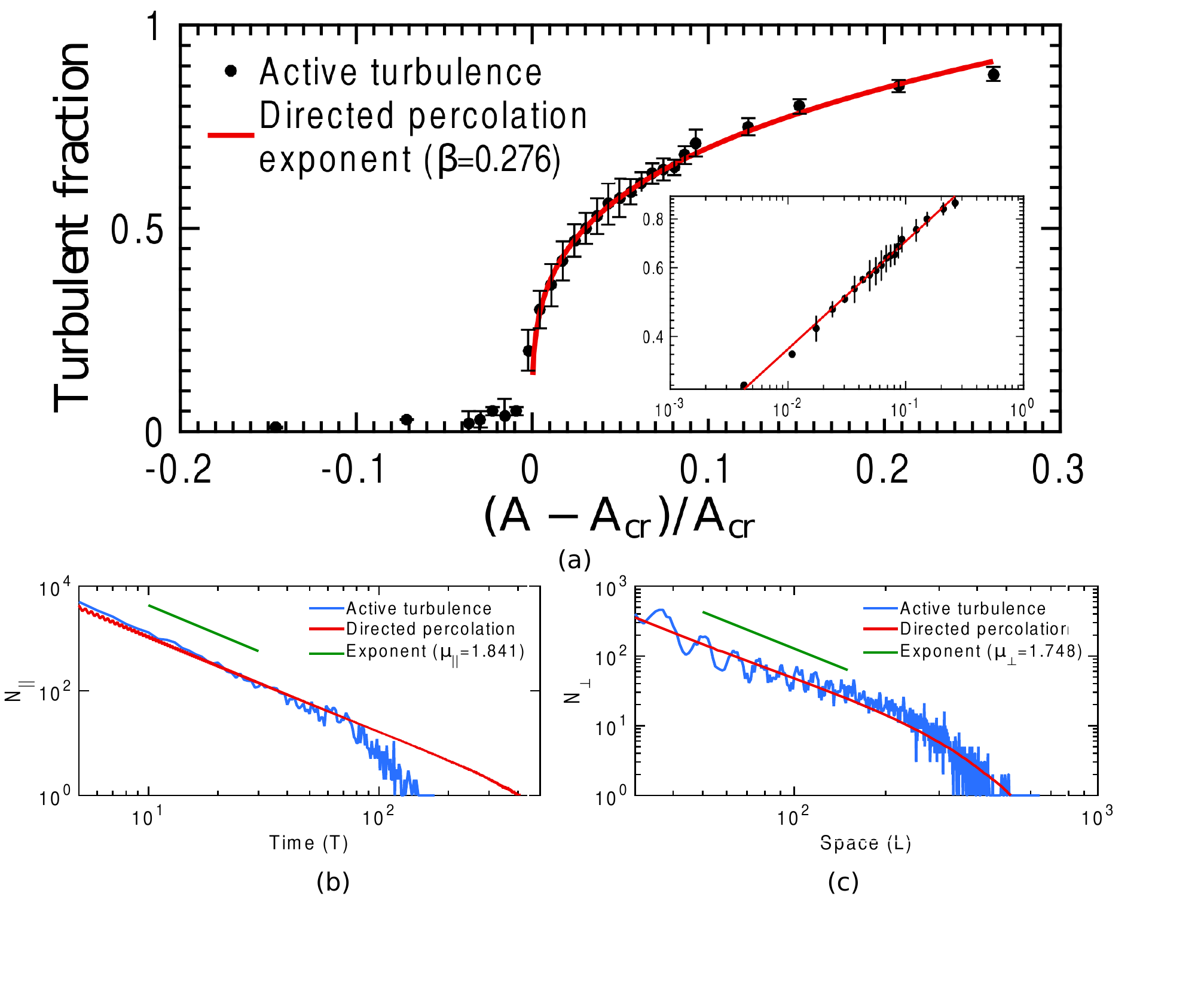}
\caption{{\bf The transition to meso-scale turbulence belongs to directed percolation universality class.} (a) Turbulence fraction as a function of the activity number. The red line corresponds to the $\text{turbulence~fraction}\propto (A-A_{c})^{\beta}$ with $\beta=0.276$ for $(1+1)$ directed percolation with spontaneous activation. Distribution of the vortex-lattice gaps is shown in (b) time and (c) space at the critical activity number. The green lines in (b) and (c) show $N_{||}\propto T^{-\mu_{||}}$, $N_{\perp}\propto L^{-\mu_{\perp}}$, respectively with $\mu_{||}=1.84,~\mu_{\perp}=1.748$ for $(1+1)$ directed percolation.}
\label{fig:turbfraction}
\end{figure}

We thus measure the turbulence fraction, the area fraction occupied by active puffs in the channel, as a function of the activity number [\fig{fig:turbfraction}(a)]. Well below the critical point, active puffs have a short lifetime and rarely split [\fig{fig:puff}(a)], leading to a negligible turbulence fraction in the steady-state [\fig{fig:turbfraction}(a)]. However, as the critical value of the activity is approached, puff decay becomes less likely and splitting time decreases substantially [\fig{fig:puff}(b)]. Above the critical point, the puff population does not die out, producing a steady-state, non-zero turbulence fraction [\fig{fig:puff}(c)], and we find the turbulence fraction continuously increases with a power-law dependence $\sim(A-A_{cr})^\beta$ [\fig{fig:turbfraction}(a)]. We measure the exponent to be $\beta=0.275 \pm 0.043$, which closely matches the universal exponent of the $(1+1)$ directed percolation process ($\beta=0.276$) and is in agreement 
with the value that has recently been measured for inertial turbulence in Couette flow ($\beta=0.28 \pm 0.03$) \cite{Hof2016}. This is striking as it draws a parallel between the zero-Reynolds number meso-scale turbulence in living fluids, which possesses a characteristic vorticity length scale, and high Reynolds number inertial turbulence, which is scale-invariant. 
Since the internal activity can spontaneously create active puffs from the absorbing vortex-lattice state with a small probability, the transition corresponds to directed percolation with spontaneous site activation, as in a weak external field \cite{Hinrichsen2000}, while the transition to inertial turbulence maps to the zero field limit (see Supplementary Information). 

To scrutinize the critical behaviour at the transition point, we further measure the spatial and temporal distributions of vortex-lattice gaps in the absorbing state (see Supplementary Information).
These distributions of the absorbing state characterise correlations of the active puffs\cite{Julia1981} and obey power laws with exponents $\mu_{\perp},~\mu_{||}$ for space and time, respectively [\fig{fig:turbfraction}(b),~(c)]. The temporal exponent is measured to be $\mu_{||}=1.84\pm0.04$ and the spatial exponent is $\mu_{\perp}=1.8 \pm 0.1$. These values also correspond to the exponents for $(1+1)$ directed percolation ($\mu_{||}=1.84,~\mu_{\perp}=1.748$)\cite{Hinrichsen2000}. The values of the critical exponents obtained from our measurements for meso-scale turbulence in a channel and for $(1+1)$ directed percolation with spontaneous site activation are summarised in Table~\ref{tab} and are compared with the experimentally measured exponents for the inertial turbulence in simple shear experiments in one dimensional geometries \cite{Hof2016}. It would not be unexpected that directed percolation universality class will continue to hold in higher dimensions as in experiments on inertial turbulence in passive liquid crystals \cite{Sano2007,Sano2009} and in channel flows \cite{Sano2015}.

Our findings present a first concrete connection between turbulence in living fluids and classical scale-invariant turbulence, beyond a superficial visual similarity, by showing that the transitions to these two profoundly distinct types of spatio-temporal disorder in channel flows belong to the same universality class, namely that the critical behaviour is represented by a directed percolation process. This opens new possibilities for further investigation of the nature of meso-scale turbulence and using tools from non-equilibrium statistical mechanics to explain critical behaviours in biological systems. Future research should investigate the transitions between ordered flow states and applicability of the directed percolation universality class in higher dimensions and in complex biological fluids.



\newpage
\section*{Methods}
\subsection*{Active nematohydrodynamics simulations}
The spatiotemporal evolution of a living fluid is described by active nematohydrodynamics equations based on the theory of liquid crystals. This formulation has been extensively applied to biological systems including bacterial suspensions \cite{Volfson2008}, microtuble/motor protein mixtures \cite{Dogic2012,Giomi2013,ourprl2013} and cellular monolayers \cite{Julicher2008,ourSM2015}. The total density $\rho$ and the velocity field ${\bf u}$ of the active matter obey the incompressible Navier-Stokes equations
\begin{align}
\bnabla\cdot\vec{u} &=0,\label{eqn:cont}\\
\rho\left(\partial_t + \vec{u}\cdot\bnabla\right) \tens{u} &= \bnabla\cdot\tens{\Pi},
\label{eqn:NS}
\end{align}
where $\tens{\Pi}$ is the stress tensor. While several studies of meso-scale turbulence have characterised the dynamics of the flow using only the velocity field as the relevant order parameter \cite{Julia2012,Jorn2013}, an additional order parameter field is required to account for the orientational order of active fluids. This is particularly important since several experiments have now established the existence and pivotal role of the orientational order in the dynamics of bacterial suspensions \cite{Volfson2008,Sano2016bacteria}, microtuble bundles \cite{Dogic2012,Francesc2016}, assemblies of fibroblast cells \cite{Silberzan2014}, and more recently in stem cell cultures \cite{Sano2016cells}. To account for the macroscopic orientational order of microscopic active and anisotropic particles, the nematic tensor $\tens{Q}=\frac{3q}{2}( \vec{n}\vec{n}-\tens{I}/3)$ is considered, where $q$ denotes the coarse-grained magnitude of the orientational order, $\vec{n}$ is the director, and $\tens{I}$ the identity tensor. The nematic tensor evolves as
\begin{align}
\left(\partial_t + \vec{u}\cdot\bnabla\right) \tens{Q} - \tens{S} &= \Gamma \tens{H},
\label{eqn:lc}
\end{align}
where $\Gamma$ is a rotational diffusivity and the co-rotation term
\begin{align}
\tens{S} = &\left(\lambda \tens{E} + \tens{\Omega}\right)\cdot\left(\tens{Q} + \frac{\tens{I}}{3}\right) + \left(\tens{Q} + \frac{\tens{I}}{3}\right) \cdot \left(\lambda \tens{E} - \tens{\Omega}\right) \nonumber\\ 
&\quad - 2 \lambda \left(\tens{Q} + \frac{\tens{I}}{3}\right)\left( \tens{Q} : \bnabla \vec{u}\right),
 \label{eqn:cor}
\end{align}
accounts for the response of the orientation field to the extensional and rotational components of the velocity gradients, as characterised by the strain rate $\tens{E}=(\bnabla\vec{u}^{T}+\bnabla\vec{u})/2$ and vorticity $\tens{\Omega}=(\bnabla\vec{u}^{T}-\bnabla\vec{u})/2$ tensors, and weighted by the tumbling parameter $\lambda$. The relaxation of the orientational order is determined by the molecular field,
\begin{align}
\tens{H} &= -\frac{\partial \mathcal{F}}{\partial \tens{Q}} + \bnabla\cdot\frac{\partial \mathcal{F}}{\partial( \bnabla\tens{Q})}, \label{eqn:molpot}
\end{align}
where $\mathcal{F}=\mathcal{F}_{b}+\mathcal{F}_{el}$ denotes the free energy. 
We use the Landau-de Gennes bulk free energy \cite{DeGennesBook},
\begin{align}
\mathcal{F}_{b} =  \frac{A}{2}\tens{Q}^2 + \frac{B}{3}\tens{Q}^3 + \frac{C}{4}\tens{Q}^4,
\end{align}
and  $\mathcal{F}_{el} = \frac{K}{2} (\bnabla\tens{Q})^2$, which describes the cost of spatial inhomogeneities in the order parameter, assuming a single elastic constant $K$. 

In addition to the viscous stress $\tens{\Pi}^\textmd{visc} = 2 \eta \tens{E}$, \eq{eqn:NS} must account for contributions to the stress $\tens{\Pi}$ from the nematic elasticity and the activity. The nematic contribution to the stress is
\begin{align}
\tens{\Pi}^{\text{elastic}}=&-P\tens{I} +2 \lambda(\tens{Q} + \tens{I}/3) (\tens{Q}:\tens{H})\nonumber\\
&\quad-\lambda \tens{H}\cdot(\tens{Q} + \frac{\tens{I}}{3})  - \lambda (\tens{Q} + \frac{\tens{I}}{3})\cdot \tens{H}\nonumber\\
&\qquad-\tens{\nabla}\tens{Q}:\frac{\partial\mathcal{F}}{\partial(\tens{\nabla}\tens{Q})} + \tens{Q}\cdot\tens{H} - \tens{H}\cdot\tens{Q},
\end{align}
which includes the pressure $P$ \cite{Berisbook}. The active contribution to the stress takes the form $\tens{\Pi}^\textmd{act} = -\zeta \tens{Q}$ \cite{Sriram2002}, such that any gradient in $\mathbf{Q}$ generates a flow field, with strength determined by the activity coefficient, $\zeta$. 

The equations of active nematohydrodynamics (\eq{eqn:cont}-\ref{eqn:lc}) are solved using a hybrid lattice Boltzmann and finite difference method \cite{Davide2007,Suzanne2011,ourpta2014}. Discrete space and time steps are chosen as unity and all quantities can be converted to physical units in a material dependent manner \cite{Cates2008, Henrich2010, ourprl2013}. 
Simulations are performed with the parameters $A=0$, $B=0.3$, $C=-0.3$, $\Gamma=0.34$, $K=0.04$, $\lambda=0.3$, $\rho=1$, and $\mu=2/3$, in lattice Boltzmann units. These specific parameters are chosen to match those previously used to quantitatively probe velocities fields of the active nematohydrodynamics of kinesin/microtuble bundles in experiments {\cite{Dogic2012}}.

We use a channel with a height $h=25$ and length $L=3000$. No-slip boundary conditions are applied to channel walls and periodic boundary conditions are used at the channel extremities. The results reported here are for homogeneous boundary conditions for the director field on the channel walls. In addition, we have performed simulations with homeotropic and weak anchoring boundary conditions and find that the transitions described in the main text are independent of the anchoring boundary conditions on the walls. 

\subsection*{Directed percolation model with spontaneous activation}

To examine the behaviour of the (1+1) directed percolation universality class, we utilized the Domany-Kinzel cellular automaton \cite{DomanyKinzel1984} and chose probabilities to correspond to bond directed percolation in the presence of a weak external conjugated field. 
We define a diagonal square lattice with empty sites corresponding to the absorbing phase (the vortex lattice state in confined active flows) and occupied sites corresponding to the activated phase (active puffs of meso-scale turbulence). 
At time $t$ each site is occupied with some probability $p_2$ if both backward sites (at time $t-1$) are occupied, with probability $p_1$ if only one backward site is occupied, and with probability $p_0$ for spontaneous site activation if neither backward site is occupied. 
Bond directed percolation in the presence of a weak external $h$ is recovered with the choice $p_2 = p_1 \left(2 - p_1\right)$ \cite{Lubeck2006} and $p_0=h\neq0$ \cite{Lubeck2002}. 
In the confined active nematic, we find that $p_0$ is small but has a non-zero value since spontaneous puff creation is observed. 

Our directed percolation simulations employ periodic boundary conditions and a lattice size of $10^4$ sites in the spatial dimension to coincide with the lattice Boltzmann system. Data is obtained from $10^3$ runs of $5\times10^3$ time steps each. 
We consider $p_0=\left\{ 0, 10^{-9}, 10^{-8}, 10^{-7}, 10^{-6} \right\}$ and find the critical probability $p=0.64470\pm 0.00002$ as expected (0.6447001(1) \cite{Jensen1996,Hinrichsen2000}). 
Comparing directed percolation with spontaneous activation simulations to the kymographs in the main text suggests that active puff creation is unlikely, and so we use $p_0 = 10^{-7}$ as a reasonable estimate. 
Measuring $N_\perp$ and $N_\parallel$, as for the lattice Boltzmann simulations, supplies the critical exponents reported in the main text. 

\subsection*{Calculating the turbulence fraction}
To obtain the active puffs, the enstrophy field $\varepsilon(x,y,t)=\vec{\Omega}\cdot\vec{\Omega}$ is calculated from the vorticity field $\vec{\Omega}(x,y,t)$. The enstrophy field is averaged across the channel $\epsilon(x,t)=\left<\varepsilon(x,y,t)\right>_y$ to obtain space-time diagrams (kymographs) of the enstrophy. As described in the main text, the channel-averaged enstrophy in the vortex-lattice phase shows regular periodic oscillations, while local active turbulence domains (the active puffs) exhibit fluctuating, noisy enstrophy signals (Fig. S1(a)).

To relate the height-averaged enstrophy signal to the onset of meso-scale turbulence, we perform image processing on the kymographs. Each kymograph is Fourier transformed in both time and space. 
The primary peaks are masked in reciprocal space-time to produce the kymographs without the structured oscillations of the periodic, background of the vortex lattice (Fig. S1(b)). 
To quantify the turbulence fraction within the channel, the existence and extent of turbulent active puffs must be automatically measured. 
Local active puffs are detected from the unmasked kymographs by dividing the height-averaged enstrophy signal into small time intervals going from $t_i$ to $t_{i+n}$. For every binned time interval, the discretized temporal autocorrelation function $c_k(x)=\frac{1}{n-1}\sum_{j=1}^{n-k}(\epsilon_{j}(x)-\bar{\epsilon})(\epsilon_{j+k}(x)-\bar{\epsilon})$ is calculated, where $k$ runs over time intervals from $0$ to $n$, $\epsilon$ is the averaged enstrophy over the discretised signal sample, and $\bar{\epsilon}$ is the enstrophy signal averaged over all time and space. 
For every $x$ point and for every binned time interval, we determine if the autocorrelation function periodic or non-periodic, indicating that the point belongs to the absorbing vortex lattice state, or an activated turbulent region, respectively. By averaging these intervals, the turbulence fraction is obtained. 

The spatial interval distribution $N_\perp$ of the absorbing state (the vortex-lattice) is measured from the processed kymographs by recording the length intervals between active puff regions for fixed temporal coordinates. Similarly, the time interval distribution $N_\parallel$ between puffs is found by recording the temporal duration of the absorbent state regions for fixed spatial coordinates. At short spatial intervals $L$, $N_\perp$ exhibits oscillations, which represent the characteristic size of the repeating vortex-lattice state.
\bibliography{refe}
\bibliographystyle{ScienceAdvances}
\noindent \textbf{Acknowledgements:} 

\noindent \textbf{Funding:} This work was supported through funding from the ERC Advanced Grant 291234 MiCE and we acknowledge EMBO funding to T.N.S (ALTF181-2013). We thank Paul Van der Schoot.\\
\newpage
\begin{table}
\begin{center}
\caption{{\bf Critical exponents for the transition to the meso-scale turbulence in a micro-channel.} Comparison to experimental measurements of inertial turbulence in Couette flow \cite{Hof2016} and directed percolation exponents \cite{Hinrichsen2000}.}
\begin{tabular}{ c c c c}
 {\bf Critical exponents} & $\beta$ & $\mu_{\perp}$ & $\mu_{||}$ \\
 \hline 
 Active turbulence at zero-Reynolds number & 0.275$\pm$0.043 & 1.80$\pm$0.10 & 1.84$\pm$0.04 \\  
 Couette experiments for inertial turbulence \cite{Hof2016} & 0.28$\pm$0.03 & 1.72$\pm$0.05 & 1.84$\pm$0.02 \\
 $(1+1)$ directed percolation \cite{Hinrichsen2000} & 0.276 & 1.748 & 1.84 \\
 \hline   
\end{tabular}
\label{tab}
\end{center}
\end{table}
\setcounter{figure}{0}
\renewcommand{\thefigure}{S.\arabic{figure}}
\begin{figure}
\centering
\subfigure[]
{\includegraphics[trim = 0 0 0 0, clip, width=0.49\textwidth]{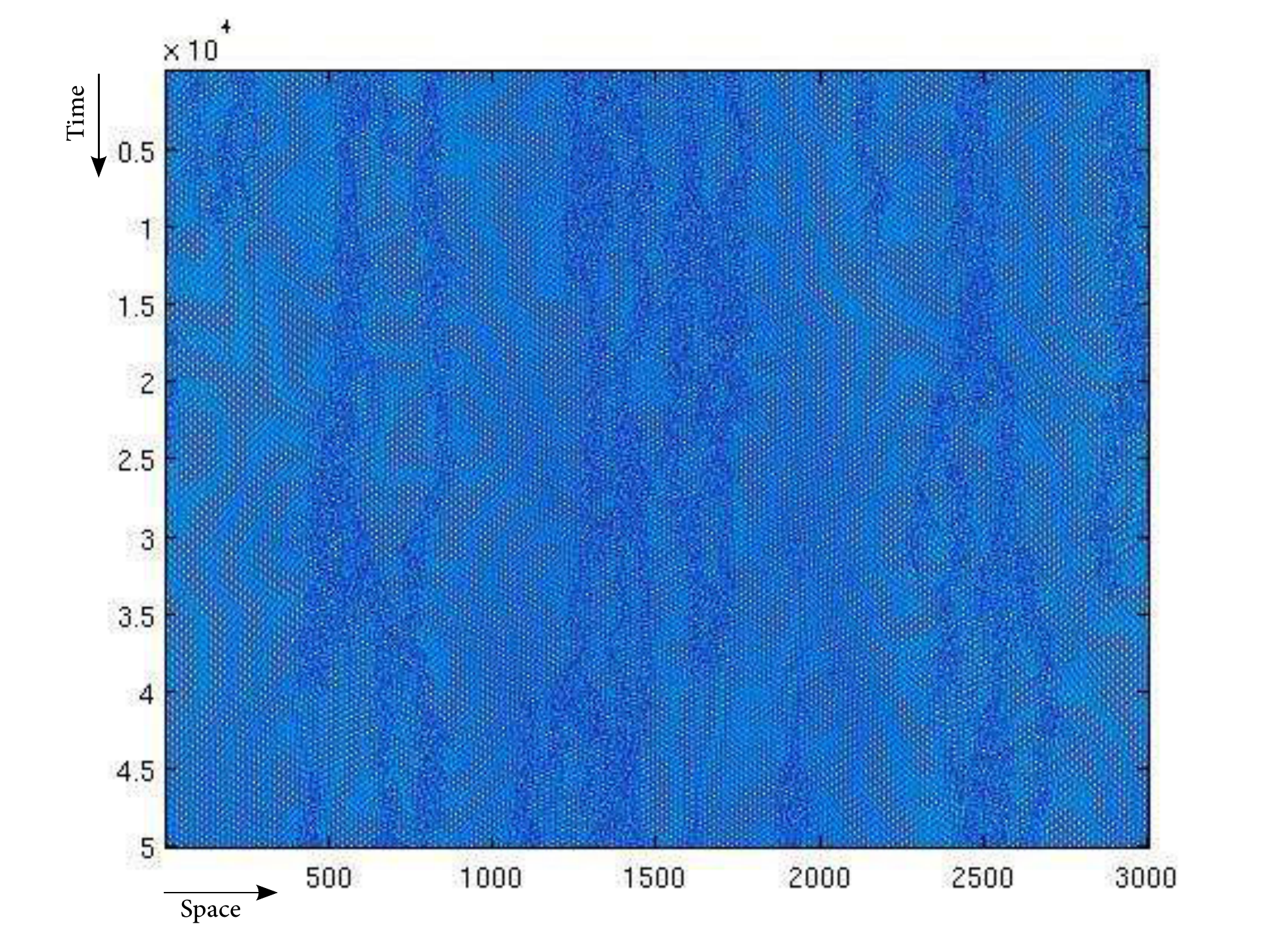}}
\subfigure[]
{\includegraphics[trim = 0 0 0 0, clip, width=0.49\textwidth]{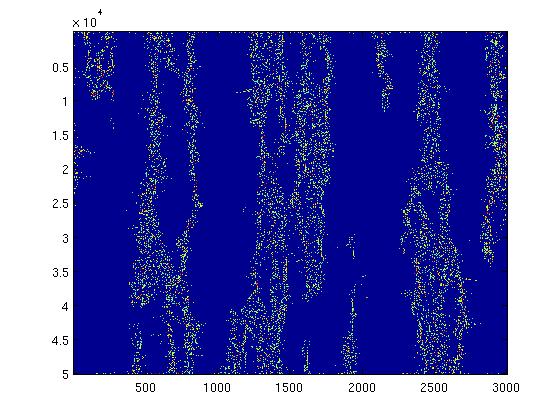}}
\caption{(a) Unmasked and (b) masked images representing a sample kymograph of height-averaged enstrophy.}
\label{fig:enst1}
\end{figure}
%


\end{document}